\documentclass[a4paper]{jpconf}
\usepackage{graphicx}
\begin{document}
\title{Towards the M(atrix) model action in an arbitrary 11D supergravity background. Progress report}

\author{Igor A. Bandos}

\address{Department of
Theoretical Physics, University of the Basque Country
 UPV/EHU,
P.O. Box 644, 48080 Bilbao, Spain and IKERBASQUE, Basque Foundation for Science, Bilbao, Spain}

\ead{igor\_bandos@ehu.es, bandos@ific.uv.es}

\begin{abstract}

 We are searching for the action principle for multiple M$0$--brane (multiple M-wave or mM0) system starting from the mM0 equations of motion obtained in the frame of superembedding approach. Surprisingly, the way from these equations to the action happens to be hampered by a problem which suggests a possible generalization of the action principle which we  call ''hierarchical action principle''.

\end{abstract}



\renewcommand{\theequation}{\arabic{equation}}

To our best knowledge, the
M$0$--brane, which can be also called M-wave or 11D massless superparticle, was described for the first time in \cite{B+T=Dpac}, where it was used to derive the action for D$0$-brane (Dirichlet particle, which is 10D type II massive superparticle) by generalized dimensional reduction. As far as the effective action for multiple M0-brane system is concerned, the construction of its purely bosonic limit  starting from the 10D ''dielectric brane action'' by Myers \cite{Myers:1999ps} was the subject of \cite{YLozano+=0207}. However, as far as the 10D Myers action does not possesses neither supersymmetry nor Lorentz invariance, it should not be surprising that the Multiple M-wave or Multiple graviton action of \cite{YLozano+=0207} is also purely bosonic and is not Lorentz invariant.

Recently the {\it supersymmetric and $SO(1,10)$ Lorentz covariant  equations of motion}  for multiple M-wave system in an arbitrary supergravity background  were derived in  \cite{mM0=PLB,mM0=PRL} in the frame of superembedding approach \cite{bpstv,hs96,Dima99,IB09:D0}. Probably these equations are 
approximate, but, if so, clearly going beyond the previously known approximations. 
The natural question is what is the action principle reproducing these set of equations? 
Surprisingly enough, construction of such an action meets a problem which we describe in this contribution. We also notice that possible resolution might consist in using a modification of the action principle which can be called ''hierarchical action principle''.

\section{Action for single $mM0$ and equations for center of energy motion of $mM0$ system}


The set of equations derived in \cite{mM0=PLB,mM0=PRL} is naturally split on the equations for the center of energy motion  and the  equations for the relative motion of the constituents of the mM0 system. The former subset of equations is obtained in \cite{mM0=PLB,mM0=PRL}  as a consequence  of the so-called superembedding equation, the basic equation of the superembedding approach to a single brane \cite{bpstv,Dima99}, and formally coincide with the equations of motion for a single M$0$-brane (11D massless superparticle). These can be obtained from the action \cite{IB07:M0} (see \cite{IB=90,IB+AN=95} for D=4 and D=3,6,10)
\begin{eqnarray}\label{M0ac=}
 S_{M0}  =  \int_{W^1} \;   \rho^{\#} \hat{E}^{=} :=   \int_{W^1} \;   \rho^{\#} d\hat{Z}^{M}{E}_{M}^{a}(\hat{Z}) u_a^{=} \; . \qquad
\end{eqnarray}
In it $W^1$ is the superparticle worldline parametrized by proper time $\tau$, $\;\hat{Z}^{M}(\tau)= (\hat{x}^\mu(\tau), \hat{\theta}{}^{\check{\alpha}}(\tau))$ are coordinate functions describing the embedding of this worldline into the target 11D superspace $\Sigma^{(11|32)}$ with local coordinates
$Z^M=(x^\mu, \theta{}^{\check{\alpha}})$ (the set of which includes eleven bosonic coordinate $x^\mu$, $\mu=0,1,\ldots, 9,10$, and $32$ fermionic coordinates ${\theta}{}^{\check{\alpha}}$, ${\check{\alpha}}=1,\ldots, 32$;  here $32$ is the dimension of the Majorana spinor representation of $SO(1,10)$),  $u_a^{=}=u_a^{=}$  is a light--like vector variable, 
$u_{ {a}}^{=} u^{ {a}\; =}=0$,
$\; \hat{E}^{=}=\hat{E}^au_a^{=}$, and  $\hat{E}^a= d\hat{Z}^{M}{E}_{M}^{a}(\hat{Z}) $ is the pull--back to $W^1$ of the bosonic supervielbein ${E}^a= d{Z}^{M}{E}_{M}^{a}({Z}) $ of the target superspace. The detailed study of the action (\ref{M0ac=}) can be found in \cite{IB07:M0} so that our presentation here will be brief and schematic.

The set of the 11D supervielbein forms 
$dZ^{ {M}}E_{ {M}}{}^{ {A}}(Z)=
 ({E}^{ {a}}, {E}^{ {\alpha}})$, 
including, in additional to the bosonic vectorial form ${E}^{a}$ ($a=0,1,\ldots, 9, 10$) also fermionic spinorial form ${E}^{ {\alpha}}= d{Z}^{M}{E}_{M}^{\alpha}({Z}) $ (${\alpha}=1,\ldots , 32$), describes  11D supergravity background when obeys the set of superspace constraints \cite{CremmerFerrara80,BrinkHowe80}. The most important of these constraints determine the bosonic torsion 2--form of  $\Sigma^{(11|32)}$. This reads
\begin{eqnarray}
\label{Ta=11D} & T^{{a}}:= DE^{{a}} =
-i{E}^\alpha \wedge  {E}^\beta \Gamma^{{a}}_{\alpha\beta}\; , \qquad
\end{eqnarray}
where $\Gamma^{{a}}_{\alpha\beta}=\Gamma^{{a}}_{\beta\alpha}$ are 11D Dirac matrices (see Appendix A), $\wedge$ denotes the exterior product of differential forms. The constraints (\ref{Ta=11D}) can be derived  from the requirement that the M$0$-brane action  (\ref{M0ac=}) is invariant under  the following local fermionic $\kappa$--symmetry transformations
\begin{eqnarray}
\label{M0kappa=}  i_\kappa \hat{E}{}^\alpha := \delta_\kappa \hat{Z}{}^M E_M^\alpha (\hat{Z})= \kappa^{+q}v^{-\alpha }_q   \; , \qquad  i_\kappa \hat{E}{}^a=0\; , \quad
 \delta_\kappa u_a^= = 0\; , \quad \delta_\kappa \rho^{\#}=0 \; . \quad
\end{eqnarray}
Here $\kappa^{+q}= \kappa^{+q}(\tau)$ is the $\kappa$--symmetry parameter, $q=1,\ldots , 16$ can be considered as $SO(9)$
spinor index and $v_q^{-\alpha}$ is the set of 16 constrained 11D spinors ($\alpha=1,\ldots, 32$)
which is related to the light--like vector $u_a^{=}$ by
\begin{eqnarray} \label{Iu--=vGv}  v_{q}^- {\Gamma}_{ {a}} v{}_{p}^- = \; u_{ {a}}{}^{=} \delta_{qp}\; , \; (a) \qquad  \delta_{qp}
u^{=}_{ a}= v_q^{-\alpha} \Gamma^a_{\alpha\beta} v_p^{-\beta}\; , \; (b) \qquad and \qquad
 \label{v-Cv-=0}   v_q^{-\alpha} C_{\alpha\beta} v_p^{-\beta}=0 \;  . \; (c) \;
\end{eqnarray}


This set of constrained spinors can be also used to write the fermionic  equations of motion of the M$0$ brane
as follows
\begin{eqnarray} \label{Ef=efv+}
\hat{E}^\alpha :=d\hat{Z}^M(\tau) E_M{}^\alpha (\hat{Z}(\tau))=
e^{+q} v_q^{-\alpha}\; . \; \qquad
\end{eqnarray}
The bosonic equations of motion can be also written in terms of the light--like vector $u_a^{=}$ and properly chosen covariant derivative. Their set includes the relations
\begin{eqnarray} \label{Ea=e++u--}
& \hat{E}^a :=d\hat{Z}^M(\tau) E_M{}^a (\hat{Z}(\tau))=
{1\over 2}e^{\#} u_a^{=}\; , \; (a) \qquad
 \label{Du--=0}
{D}u_a^{=}=0 \; , \; (b) \qquad Dv^{-\alpha}_q=0 \; , \; (c) \; , \qquad
\end{eqnarray}
where $D$ is an appropriate covariant derivative. In our worldline approach $D= d\tau D_\tau =e^\# D_\#$,  where  $e^\#=d\tau e_\tau^\#(\tau)$ is the einbein form, but in the superembedding formalism of \cite{mM0=PLB,mM0=PRL} it has also fermionic components ($D=e^\# D_\# + e^{+q} D_{+q}$) while (\ref{Du--=0}) still holds. These covariant derivative enters as well in the equation for the Lagrange multiplies $\rho^{\#}$ which reads
\begin{eqnarray} \label{Drho++=0}
{D}\rho^{\#}:=d \rho^{\#} - 2 \rho^{\#}\omega^{(0)}=0\; , \qquad
\end{eqnarray}
where $\omega^{(0)}$ is a certain $SO(1,1)$ connection and, as above,  $D=e^{\#}D_{\#}$.

To define the $SO(1,10)\times SO(1,1)\times SO(9)$ covariant derivative used in  (\ref{Du--=0}) and below it is convenient, to complete the light--like vector $u_a^{=}$ till the full {\it moving frame}  containing, besides $u_{{a}}^{=}(\tau)$, also a complementary light-like vector  $u_{{a}}^{\#}(\tau)$ and $9$ normalized spacelike  vectors  $u_{{b}}^i(\tau)$ orthogonal to both of them and among themselves, 
\begin{eqnarray}\label{u++u++=0=}
u_{ {a}}^{=} u^{ {a}\; =}=0\; , \qquad u_{ {a}}^{\# } u^{ {a}\; \#
}=0\; , \qquad    u_{ {a}}^{\; \# } u^{ {a}\; =}= 2\; , \quad
\nonumber \\  \label{uiuj=-}  u_{ {a}}^{=} u^{ {a}\,i}=0\; , \qquad
 u_{{a}}^{\;\#} u^{ {a} i}=0\; , \qquad   u_{ {a}}^{ i} u^{ {a} j}=-\delta^{ij}\; . \quad
\end{eqnarray}
With this vectors one can split Eq. (\ref{Ea=e++u--}a) on
\begin{eqnarray} \label{E--=0}
\hat{E}^{=}:= \hat{E}^a u_a^{=}=0 \; , \quad (a) \qquad  \label{Ei=0}
\hat{E}^{i}:= \hat{E}^a u_a^{i}=0 \; , \quad (b) \qquad
 \label{Eu++=e++}
and \qquad \hat{E}^{\#}:= \hat{E}^a u_a^{\#}=e^{\#} \; . \; (c)
\end{eqnarray}
The latter equation defines the einbein induced by embedding of $W^1$ into $\Sigma^{(11|32)}$.
The fermionic equations of motion for a single M$0$ brane, and also for the center of mass of the mM$0$ system in the superembedding description of \cite{mM0=PLB,mM0=PRL}, can be written in the form of
\begin{eqnarray} \label{E-q=0}
\hat{E}_q^{-}:= \hat{E}^\alpha v_{\alpha q}{}^{-}= 0 \; . \; \qquad
\end{eqnarray}
where $v_{{\alpha}q}^{\; -}$ is related to above  $v_{q}^{- \beta }$ by $v_{\alpha}{}^{-}_q =  i
C_{\alpha\beta}v_{q}^{- \beta }$. The equivalence of Eqs. (\ref{E-q=0}) and (\ref{Ef=efv+}) holds due to the constraint (\ref{v-Cv-=0}c). Further details on the worldline geometry induced by this embedding, on the $SO(1,10)\times SO(1,1)\times SO(9)$ covariant derivative $D$  as well as on moving frame and spinor moving  frame variables can be found in \cite{IB07:M0,mM0=PLB}.

The above equations describe as well the center of energy motion of the mM$0$ system as described by superembedding approach of \cite{mM0=PLB,mM0=PRL}.

\section{Lagrangian which might be used to produce the equation of relative motion of the $mM0$ constituents}

The relative motion of the constituents of the system of $N$ M0--branes was described in \cite{mM0=PLB,mM0=PRL} in terms of the nanoplet of anti-hermitian bosonic $N\times N$ matrix fields ${\bf X}^i$
with $SO(1,10)$ weight $2$, the fact which can be expressed by writing ${\bf X}^i={\bf X}^i_{\#}={\bf X}^i_{++} $, by a 16-plet of fermionic $N\times N$ matrix fields ${\bf \Psi}_{q}$
with $SO(1,1)$ weight 3, $\; {\bf \Psi}_{q}={\bf \Psi}_{\# \, +q}={\bf \Psi}_{++\, +q}$,
\begin{eqnarray}\label{PsiinSU}
& {\bf X}^i (\tau) \; \in \; su(N) \; , \qquad i=1,\ldots, 9\, , \qquad {\bf \Psi}_{q}(\tau) \; \in \; su(N) \; , \qquad  q=1,\ldots, 16\, ,
\end{eqnarray}
and by (auxiliary; actually, pure gauge) one dimensional $SU(N)$ connection ${\bf A}_{\#}$.   
These fields are inert under the 11D Lorentz $SO(1,10)$ transformations. The equations for ${\bf X}^i (\tau)$ and ${\bf \Psi}_{q}(\tau)$ obtained in \cite{mM0=PRL} involve the projections $\hat{F}_{\# ijk}:= F^{{a}{b}{c}{d}}  u_{{a}}
{}^{=}u_{{b}}{}^{i}u_{{c}}{}^{j}u_{{d}}{}^{k}$, $\hat{T}_{\#\, i\, +q} :=T_{{a}{b}}{}^{{\beta}}(\hat{Z}) v_{{\beta}q}^{\; -}\,
u_{{a}}^{=}u_{{b}}^i\; $, $\hat{R}{}^{ =i}:=  \hat{R}^{ {a} {b}}u_{ {a}}^{=} u_{ {b}}{}^i=0$ of the four form flux, gravitino field strength and curvature tensor;
they  can be reproduced by varying Lagrangian
\begin{eqnarray}\label{L+8=}
& {\cal L} := {\cal L}_{\#\#\#\#} = tr \left( {1\over 2}D_{\#} {\bf X}^i D_{\#} {\bf X}^i - {1\over 64} ([{\bf X}^i ,{\bf X}^j ])^2  - 4i {\bf \Psi}_q D_{\#} {\bf \Psi}_q +2i {\bf X}^i {\bf \Psi}_q \gamma^i_{qp} {\bf \Psi}_p \right) +  \quad \nonumber \\ &  + {1\over 2} \hat{R}_{\# i\# j} \;  tr  ({\bf X}^i{\bf X}^j) + {1\over 12} \hat{F}_{\# ijk} \;  tr  ({\bf X}^i{\bf X}^j{\bf X}^k)
+ {i\over 3} \hat{F}_{\# ijk} \;  tr  ({\bf \Psi}\gamma^{ijk}{\bf \Psi})-2i \hat{T}_{\# i\; +q} \;  tr  ({\bf X}^i{\bf \Psi}_q)
\end{eqnarray}
with respect to $\Psi_q$, $A_\#$ and ${\bf X}{}^i$, and {\it omitting the complete derivatives}.
This corresponds to the variation of action $S_{relative}= \int \mu^{\#\#\#\#}  {\cal L}_{\#\#\#\#}$ constructed by integrating the Lagrangian (\ref{L+8=}) with a proper measure $\mu^{\#\#\#\#}$  allowing integration by parts,
and omitting the boundary terms.

\section{Problems with the measure for the mM$0$ action and possible relevance of the  ''hierarchic action principle''}

The natural candidates for the measure with the appropriate $SO(1,1)$ weight ($-8$) are
 \footnote{Notice that in our notation the positive weight 1 with respect to $SO(1,1)$ group is denoted either by subindex $_+$  or by a superindex $^-$, so that one can write, for instance, $\; {\bf \Psi}_{q}={\bf \Psi}_{\# \, +q}={\bf \Psi}_{\#}{}^-_{q}= {\bf \Psi}{}^{=\, - q}$, $\;\rho^{\#}= \rho^{++}=\rho_{--}=\rho_{=}$.  }  $ \hat{E}^{\#}  (\rho^{\#})^3$ and
$ \hat{E}^{=}  (\rho^{\#})^5$. However, the first one, when used in the
$S_{mM0}=S_{M0}+ S_{relative}$, would result in breaking of a number of gauge symmetries characteristics of the center of energy action $S_{M0}$. This forces us to consider the other candidate, $\mu^{\#\#\#\#} =\hat{E}^{=}  (\rho^{\#})^5$, with which
 \begin{eqnarray}\label{Srel=}
 S_{relative}  =  \int_{W^1} \;   \hat{E}^{=}  (\rho^{\#})^5 {\cal L}_{\#\#\#\#}  \; .   \qquad
\end{eqnarray}
The variation of (\ref{Srel=}) does reproduce the equations of relative motion of the mM$0$ constituents as they were obtained in \cite{mM0=PRL}, but multiplied by $\hat{E}_\tau^{=}$. Then the problem is that $\hat{E}_\tau^{=}$ vanishes on the ''center of energy mass shell'' due to Eq. (\ref{E--=0}a), $\hat{E}^{=}=d\tau^{=}\hat{E}_\tau^{=}=0$. Notice also that, when this equation is taken into account, the functional (\ref{Srel=}) vanishes itself so that the variational problem for $S_{relative} $ does not make sense on the center of energy mass shell.

These simple observations suggest the following (probably, looking a bit artificial) 
prescription of extracting the equations of \cite{mM0=PLB,mM0=PRL}  from (\ref{Srel=}).
First one does not set $\hat{E}_\tau^{=}$ equal to zero and vary the action (\ref{Srel=}) with respect to the relative motion variables ${\bf \Psi}_q$, ${\bf A}_\#$ and ${\bf X}^i$. Then one divide the equations thus obtained by  $\hat{E}_\tau^{=}$ (thus compensating the $\hat{E}_\tau^{=}$ multipliers which are present in them) and, only after that, one uses the center of energy equations, including Eq. (\ref{E--=0}a), $\hat{E}_\tau^{=}=0$. To resume, the equations for relative motion of  mM$0$ constituents can be obtained as
\begin{eqnarray}\label{fEq=vSrel}
& \lim\limits_{\hat{E}_\tau^{=}\mapsto 0} {1\over \hat{E}_\tau^{=}}{\delta S_{relative}\over \delta {\bf \Psi}_q} =0\; , \qquad
\lim\limits_{\hat{E}_\tau^{=}\mapsto 0} {1\over \hat{E}_\tau^{=}}{\delta S_{relative}\over \delta {\bf A}_\#} =0\; , \qquad  \lim\limits_{\hat{E}_\tau^{=}\mapsto 0} {1\over \hat{E}_\tau^{=}}{\delta S_{relative}\over \delta {\bf X}{}^i} =0\; . \qquad
\end{eqnarray}

Such a prescription of 'hierarchical action principle', formulated above for the relative motion, can be extended to the  complete description of the multiple M$0$ system by the action
\begin{eqnarray}\label{mM0ac=}
 S_{mM0}  = S_{M0} + S_{relative}  =  \int_{W^1} \;    \hat{E}^{=} \rho^{\#} \left( 1+
 (\rho^{\#})^4 {\cal L}_{\#\#\#\#}\right)   \; , \qquad
\end{eqnarray}
where ${\cal L}_{\#\#\#\#}$ is defined in (\ref{L+8=}). Variation of (\ref{mM0ac=}) produces all the equations of center of energy motion of the mM$0$ system, including Eq. (\ref{E--=0}a) (when one assumes a generic situation with $ {\cal L}_{\#\#\#\#}\not= - {1\over 5(\rho^{\#})^4}$). Then the reason for the special r\^ole of Eq. (\ref{E--=0}a) is that the action  (\ref{mM0ac=}) vanishes due to it, and  the equations of the relative motion are obtained as in (\ref{fEq=vSrel}).

The selfconsistency of the above described {\it ''hierarchical action principle''} requires additional study (hence the 'progress report' in the title).
Notice that some generalizations of the action principle were proposed and used before. Let us mention  the generalized action principle of the group manifold approach to supergravity \cite{rheonomic}, its p-brane counterpart \cite{bsv95} and the  'democratic action' \cite{democratic} for 10D  type II supergravity\footnote{An example of a more radical modification is the two-time `multiaction' construction of \cite{Manvelyan:2001pa}.}.

{\bf Aknowledgments}
The author is thankful to Dmitri Sorokin
for useful discussion.
The partial support from the research grants FIS2008-1980 from the
Spanish MICINN  and the Basque Government Research Group Grant ITT559-10 is greatly acknowledged.



\noindent
{\bf References}
\begin{thebibliography}{99}
\renewcommand{\theequation}{R.\arabic{equation}}

\bibitem{B+T=Dpac}
Bergshoeff E and Townsend P K 1997 Super-$D$-branes
{\it Nucl.Phys.} {\bf B490}, 145--162. 
({\it Preprint} hep-th/9611173).


\bibitem{Myers:1999ps}
  Myers R C 1999 Dielectric-branes
  JHEP 12(1999)022
   ({\it Preprint} hep-th/9910053).



\bibitem{YLozano+=0207}
Janssen B and Lozano Y 2002 
  {\it Nucl.\ Phys.} {\bf B643}, 399;
  {\bf B658}, 281 (2003)
  ({\it Preprint} hep-th/0207199).


\bibitem{mM0=PLB}
  Bandos I A 2010a 
  {\it Phys.\ Lett.}  {\bf B687}, 258--263 
  ({\it Preprint}  arXiv:0912.5125 [hep-th]).

\bibitem{mM0=PRL}
Bandos I A 2010b
{\it Phys. Rev. Lett.} {\bf 105}, 071602 [1-4] (2010); 
  Phys. Rev. {\bf D82}, 105030 [1-19] (2010). 


\bibitem{bpstv}
   Bandos I, Sorokin D, Tonin, Pasti P and Volkov D 1995
{\it   Nucl. Phys.} {\bf B446}, 79-118 ({\it Preprint} hep-th/9501113).



\bibitem{hs96}
Howe P S and Sezgin E 1996 
 {\it  Phys.\ Lett.}  {\bf B390}, 133; {\bf B394}, 62 ({\it Preprints} hep-th/9607227, hep-th/9611008).

\bibitem{Dima99}
  Sorokin D P 2000 Superbranes and superembeddings,
  {\it  Phys.\ Rept.}  {\bf 329}, 1-101 ({\it Preprint} hep-th/9906142).

\bibitem{IB09:D0}
  Bandos I A 2009
  {\it Phys.\ Lett.} {\bf B680}, 267--273 (2009) ({\it Preprint}
  0907.4681 [hep-th]).



\bibitem{IB07:M0}
  Bandos I A 2008
{\it   Phys.\ Lett.}  B {\bf 659}, 388
 {\it  Nucl.\ Phys.}  {\bf B796}, 360
({\it Preprint}  0707.2336, 0710.4342 [hep-th]).

\bibitem{IB=90}
Bandos I A 1990  
  Sov.\ J.\ Nucl.\ Phys.\  {\bf 51}, 906--914 (1990).
\bibitem{IB+AN=95}
Bandos I A  and  Nurmagambetov A Yu 1995 
  Class. Quant. Grav.  {\bf 12}, 1881-1892 
  ({\it Preprint} hep-th/9502143).


\bibitem{CremmerFerrara80}
Cremmer E and Ferrara S 1980
{\it  Phys. Lett.} B {\bf 91}, 61.
\bibitem{BrinkHowe80}
Brink L and Howe P S,
{\it Phys. Lett.} B {\bf 91}, 384.





\bibitem{Witten:1995im}
Witten E 1996
  {\it Nucl. Phys. }   B {\bf 460}, 335 ({\it Preprint} hep-th/9510135).


\bibitem{Banks:1996vh}
  Banks T, Fischler W, Shenker S H and Susskind L 1997,
  {\it Phys.\ Rev.\ } D {\bf 55}, 5112-5128.

\bibitem{rheonomic}
 Ne'eman Y and Regge T 1978 
  {\it Phys.\ Lett.}  {\bf B74}, 54; 
  {\it Riv.\ Nuovo Cim.} {\bf 1N5}, 1.


\bibitem{bsv95}
 Bandos I A, Sorokin D P and Volkov D V 1995 
  {\it Phys.\ Lett.}  {\bf B352}, 269--275
  ({\it Preprint} hep-th/9502141).
\bibitem{democratic}
Bergshoeff E, Kallosh R, Ortin T, Roest D and Van Proeyen A 2001 
  {\it Class.\ Quant.\ Grav.} {\bf 18}, 3359.

\bibitem{Manvelyan:2001pa}
  Manvelyan R and Mkrtchyan R 2002 
  {\it Mod.\ Phys.\ Lett.}  {\bf A17}, 1393 
  ({\it Preprint} hep-th/0112233).


\end{thebibliography}
\end{document}